\documentstyle[12pt,epsf]{article}

\font\blackboard=msbm10 at 12pt
\font\blackboards=msbm7
\font\blackboardss=msbm5
\newfam\black
\textfont\black=\blackboard
\scriptfont\black=\blackboards
\scriptscriptfont\black=\blackboardss

\newcommand{\junk}[1]{}

\newcommand{\ba}{\begin{array}}
\newcommand{\ea}{\end{array}}
\newcommand{\be}{\begin{equation}}
\newcommand{\ee}{\end{equation}}
\newcommand{\bea}{\begin{eqnarray}}
\newcommand{\eea}{\end{eqnarray}}
\newcommand{\beas}{\begin{eqnarray*}}
\newcommand{\eeas}{\end{eqnarray*}}

\def\identity{{\rlap{1} \hskip 1.6pt \hbox{1}}}

\def\laplace{{\kern1pt\vbox{\hrule height 1.2pt\hbox{\vrule width
1.2pt\hskip
  3pt\vbox{\vskip 6pt}\hskip 3pt\vrule width 0.6pt}\hrule height
  0.6pt}
  \kern1pt}}
\def\scriptlap{{\kern1pt\vbox{\hrule height 0.8pt\hbox{\vrule width
  0.8pt
  \hskip2pt\vbox{\vskip 4pt}\hskip 2pt\vrule width 0.4pt}\hrule height
  0.4pt}
  \kern1pt}}

\def\roughly#1{\raise.3ex\hbox{$#1$\kern-
.75em\lower1ex\hbox{$\sim$}}}

\def\tr{{\rm tr} \,}
\def\Tr{{\rm Tr} \,}

\newcommand{\gone}[1]{}

\begin{document}
\pagestyle{plain}
\setcounter{page}{1}

\baselineskip16pt

\begin{titlepage}

\begin{flushright}
SU-ITP-01-41\\
hep-th/0110093
\end{flushright}
\vspace{8 mm}

\begin{center}

{\Large \bf The Meaning of Infrared Singularities\\
in Noncommutative Gauge Theories\\}

\end{center}

\vspace{7 mm}

\begin{center}

Mark Van Raamsdonk

\vspace{3mm}
{\small \sl Institute for Theoretical Physics} \\
{\small \sl Stanford University} \\
{\small \sl Stanford, CA, 94306 U.S.A} \\
{\small \tt mav@itp.stanford.edu}\\

\end{center}

\vspace{8 mm}

\begin{abstract}

We point out that the leading infrared singular terms in the effective actions of 
noncommutative gauge theories arising from nonplanar loop diagrams have a 
natural interpretation in terms of the matrix model (operator) formulation of 
these theories. In this formulation (for maximal spatial noncommutativity), 
noncommutative space arises as a configuration of an infinite number of  
D-particles. We show that the IR singular terms correspond to
instantaneous linear potentials between these D-particles resulting
from the zero point energies of fluctuations about this background. 
For theories with fewer fermionic than bosonic degrees of freedom, such as 
pure 
noncommutative gauge theory, the potential is attractive and renders
the theory unstable. With more fermionic than bosonic degrees of freedom, the 
potential is repulsive and we argue that the theory is stable, though
oddly behaved.        

\end{abstract}

\vspace{0.7cm}
\begin{flushleft}
October 2001
\end{flushleft}
\end{titlepage}
\newpage

\section{Introduction}

In the context of string theory, noncommutative field theories \cite{cds,dh,sw} 
appear
as a description of the low energy physics of D-branes in the
presence of constant $B$ (NS-NS two-form) fields, or equivalently, D-branes  
carrying uniform lower-dimensional brane charges. Alternatively, we
may describe the situation directly in terms of these lower dimensional
branes carrying the higher dimensional brane charge. For example, a 
noncommutative D2-brane with a uniform
zero-brane charge density may be described as a collection of an
infinite number of D0-branes whose (infinite dimensional)
configuration matrices realize the algebra of the noncommutative
plane, $[X^1,X^2] = i \theta$. The matrices describing the
fluctuations about this background may be expanded in a basis
generated by the background matrices (noncommutative coordinates)
$X^1$ and $X^2$; in other words they are fields on the noncommutative
plane. Thus, noncommutative field theory is the natural description
arising in terms of the lower
dimensional brane degrees of freedom.\footnote{The Seiberg-Witten map \cite{sw}
between noncommutative and commutative descriptions may be understood
as the map between the lower dimensional brane degrees of freedom and
those of the higher dimensional brane.}

While noncommutative field theories outside the context of string
theory are not obviously related to D-branes, there typically exists a
matrix model description (the operator formalism) analogous to the
lower dimensional brane picture in string theory. In particular, for
pure noncommutative gauge theory,
the relevant matrix model is just a bosonic version of the D0-brane
matrix model with the number of scalars equal to the number of
noncommutative dimensions. Thus, even outside the context of string
theory one may think of noncommutative gauge theory as describing
fluctuations about a particular state of an infinite number of (bosonic) 
D-particles. In this paper, we show that certain intriguing features of quantum
noncommutative gauge theories can be understood naturally in terms of
this D-particle picture. 

Perhaps the most surprising feature of quantum noncommutative field
theories is the phenomenon of UV-IR mixing. As demonstrated in
\cite{msv} for noncommutative scalar theories, high momentum virtual
particles running in loops of nonplanar diagrams can lead to long
range correlations. In particular, even for a massive theory,
nonplanar diagrams give contributions to the effective action which
exhibit infrared singularities. As demonstrated in \cite{hayakawa,
mst}, these infrared singular terms appear also in the effective
action of noncommutative gauge theories. For example, in four dimensions,
nonplanar one loop diagrams contribute a term proportional to
\be
\label{AA}
g^2 A_\mu(p) A_\nu(-p)  {\tilde{p}^\mu \tilde{p}^\nu  \over |\tilde{p}|^4} \; .
\ee
where $\tilde{p}^\mu = \theta^{\mu \nu} p_\nu$.
This term is singular as $p \to 0$ and will clearly have dramatic
consequences for the long distance physics of the theory. 

The main result of this note is to demonstrate that the term (\ref{AA})
corresponds to physics that is actually quite familiar when viewed in
terms of the D-particle picture. Indeed, it is simply an instantaneous linear
potential between the D-particles making up the noncommutative
space. If not for supersymmetry, such a potential would exist between
D0-branes in string theory due to the zero-point energies of the
strings connecting them. The famous $v^4/r^7$ potential between widely
separated D0-branes in the BFSS matrix model \cite{bfss} arises only after 
cancellation
between contributions from off-diagonal bosonic degrees of freedom and
contributions from off diagonal fermionic degrees of freedom, both of
which have linear potentials at leading order \cite{dkps}. Thus, an 
instantaneous
linear potential between nonsupersymmetric D-particles is completely
expected, with an
attractive sign for theories with more bosonic than fermionic degrees
of freedom and with a repulsive sign for theories with an excess of
fermionic degrees of freedom.
 
In the attractive case, we show that the one loop potential is
unbounded below (relative to the energy of the chosen vacuum) and
renders the theory unstable. Based on the D-particle intuition, we
expect this instability to persist nonperturbatively. Thus it appears
that pure noncommutative gauge theory is not a sensible quantum field theory.
On the other hand, with an excess of fermionic degrees of freedom, the
D-particle repulsion leads to a positive definite potential and a
``confinement'' of density fluctuations, so the theory appears to be
stable (with appropriate boundary conditions at infinity).  

The plan for the remainder of the paper is as follows. In section 2, we
 write down the leading infrared singular quadratic term in the one-loop
effective action for d-dimensional noncommutative gauge theory. This
term is gauge invariant at leading order in the noncommutativity
parameter but not under the full noncommutative gauge
transformation. In section 3, we suggest a minimal gauge invariant
completion of this leading quadratic term, involving a two point
function of the simplest gauge invariant open Wilson line
operator. In section 4, we briefly review the matrix model
formulation of the theory and then recast our proposed gauge
invariant effective action in terms of this D-particle picture in
section 5, showing that it is exactly an instantaneous linear
potential between the D-particles. A direct derivation of this potential from 
the 
matrix model is included in an appendix. In section 6, we discuss the
physical consequences of our observation, arguing that pure
noncommutative gauge theory (and theories with fewer bosonic than
fermionic degrees of freedom) is unstable due to the attractive
linear potential, while theories with an excess of fermionic degrees
of freedom are stable. A few concluding remarks are offered in section 7.

The literature on noncommutative field theories is by now very large. For recent 
reviews of the subject, including large lists of references, we refer the 
reader to \cite{harvey, dn, szabo}. For an interesting previous
discussion of infrared singularities and other properties of
noncommutative field theories
in terms of the matrix model formalism, see \cite{ikk}.

\section{Nonplanar effective action in noncommutative gauge theory}

In this section, we write down the infrared singular quadratic terms in
the
effective action of d-dimensional noncommutative gauge theory arising from
non-planar diagrams. These terms were originally calculated for 4
dimensional gauge theory in \cite{hayakawa,mst}. 

We consider the action for a $U(N)$ gauge field in d-dimensional Minkowski
spacetime, with fermions $\psi_j$ and scalars $\phi_i$, all in the adjoint 
representation,\footnote{We ignore the possibility of fundamental matter fields 
since these only give rise to planar diagrams.} 
\[
S = \int d^d x \; \tr(-{1 \over 4} F_{\mu \nu}  F^{\mu \nu} - {1
\over 2} D_\mu \phi_i D^\mu \phi_i
- {1 \over 2} m_i^2 \phi_i^2 - i \bar{\psi_j} (\gamma^\mu D_\mu + m_j)\psi_j ) 
\] 
where $F_{\mu \nu} = \partial_\mu A_\nu - \partial_\nu A_\mu + ig [A_\mu, 
A_\nu]_\star$
and $D_\mu = \partial_\mu + ig[A_\mu, \; ]_\star$. The action is invariant under a 
noncommutative gauge transformation
\be
\label{gauge}
\delta A_\mu = \partial_\mu \Lambda + ig[A_\mu, \Lambda]_\star \qquad
\delta \phi = -ig[\phi, \Lambda]_\star \qquad \delta \psi = -ig[\psi, \Lambda]_\star
\; .
\ee
It is straightforward to compute
the quadratic terms in the one-loop, 1PI effective action for the gauge field,
and it was found \cite{hayakawa, mst} that nonplanar diagrams give rise to new 
infrared singular terms of the form
\[
\int d^4 p \; \tr(A_\mu(p)) \tr(A_\nu(-p)) {\tilde{p}^\mu \tilde{p}^\nu \over 
|\tilde{p}|^4}
\]
in the four dimensional case, where $\tilde{p}^\mu = \theta^{\mu \nu} p_\nu$. 
In appendix A, we find that in d-dimensions the nonplanar quadratic effective 
action is 
\be
\label{ea}
\Gamma_{1PI} = \int {d^d p \over (2 \pi)^d} \tr(A_\mu(p))
 \tr(A_\nu(-p)){\tilde{p}^\mu \tilde{p}^\nu \over |\tilde{p}|^d}
\left\{  {g^2 \over 2 \pi^{d \over 2} } \Gamma({d \over 2}) (d - 2 +
N_s - {1 \over 2} {\cal N}_f) \right\}   
\ee
where $N_s$ and ${\cal N}_f$ are the number of scalars and the number of 
fermionic degrees of freedom respectively. 
As pointed out in \cite{msv,mst}, the leading infrared singularities cancel if 
there are an equal number of bosonic and fermionic degrees of freedom. With an 
excess of bosonic degrees of freedom, we find that the nonplanar effective 
potential ($V_{NP}$ where $\Gamma_{NP} = \int -V_{NP}$) is negative, suggesting 
tachyonic behavior for the low momentum modes 
of the gauge field. For an excess of fermionic degrees of freedom, the effective 
potential is positive, so the infrared singular term does not appear to cause 
any instability. This behavior is opposite to that which was suggested 
previously in \cite{mst}, due to a discrepancy in the sign of the term 
(\ref{ea}).\footnote{A discrepancy with \cite{mst} has been found
previously by other authors, including \cite{llt}.} Shortly, we will
 develop an intuitive understanding of the origin of 
this term and present an independent argument that the present sign is correct. 
We leave further discussion of the consequences of this term until then.

\section{Gauge invariant effective action}

It is easy to check that the quadratic term we have calculated is invariant 
under a noncommutative gauge transformation (\ref{gauge}) to leading order in 
$\theta$, since upon the variation 
$\delta A_\mu = i p_\mu \Lambda$, our effective action cancels due to the 
vanishing 
of $p_\mu \tilde{p}^\mu = \theta^{\mu \nu} p_\mu p_\nu$. On the other hand, it is not 
invariant under the full noncommutative gauge transformation. Since the 
nonlinear term in the gauge transformation mixes terms with different numbers of 
occurrences of the gauge field, it is reasonable to expect that terms in the 
effective action at higher orders in the gauge field will complete (\ref{ea}) 
into a gauge invariant expression. This was argued to be the case for the 
effective action of the $3+1$ dimensional noncommutative ${\cal N} =
4$ theory on its Coulomb branch in 
\cite{lm,liu}. 

Since matrix products always occur together with star products in the original 
action, the double trace structure of (\ref{ea}) suggests that the one-loop 
nonplanar gauge invariant effective action should take the form 
\be
\label{wact}
\int d^dp \sum_i W_i(p) \tilde{W}_i(-p) \Delta_i(p)
\ee
where $\Delta_i(p)$ are some functions of momenta and $W_i$ and $\tilde{W}_i$ 
are operators that are separately gauge invariant, possibly with some indices 
contracted with indices on $\Delta_i$. Gauge invariant operators with non-zero 
momenta in noncommutative gauge theory generically include open Wilson lines 
\cite{iikk}, 
and take the form 
\be
\label{wilson}
W(p) = \int d^dx e^{i p \cdot x} \tr( {\cal O} \star {\rm P}_\star e^{ig 
\int_x^{x+\tilde{p}} A})
\ee
where the path-ordered Wilson line runs over the straight line path
from x to $x+\tilde{p}$ and 
${\cal O}$ is some gauge covariant operator built from field strengths and 
covariant derivatives.\footnote{More generally, we may insert a series of 
covariant operators at various points along the Wilson line and consider a 
Wilson line of arbitrary shape whose net displacement is $\tilde{p}$.} The 
simplest such operator has ${\cal O} = 1$, and has an expansion in powers of $A$ 
given by 
\bea
\omega(p) &=& \int d^dx e^{i p \cdot x} \tr({\rm P}_\star e^{i g
\int_x^{x+\tilde{p}} A} 
) \nonumber \\
&=&  (2 \pi)^d N \delta(p) + ig \; \tr(A_\mu (p)) \tilde{p}^\mu + {\cal O} (A^2)
\label{expand}
\eea
Note that the term linear in $A$ has exactly the structure appearing in the 
quadratic effective action. In particular, a term 
\be
\label{density}
\int d^dp \; \omega(p) \omega(-p) {1 \over |\tilde{p}|^d}
\ee
is gauge invariant and has a leading term with precisely the form 
(\ref{ea}).\footnote{The leading delta function term in $\omega$ may be ignored 
since it only contributes to an infinite constant term in the
potential.} The only other operator of the form (\ref{wilson}) that yields
a term linear in $A$ with the same structure is the one with ${\cal O} =
\theta^{\mu \nu} F_{\mu \nu}$. Calling this operator $A(p)$, one could also
write terms
\[
\int d^dp A(p) \omega(-p) {1 \over |\tilde{p}|^d} \; , \qquad \int d^dp A(p) A(-p) 
{1 \over |\tilde{p}|^d}
\]
that reproduce the structure (\ref{ea}). However, both of these
produce cubic terms in the effective action (coming from the nonlinear
term in $F$) that are more singular than the cubic terms obtained by a
direct calculation. Thus, it seems that the quadratic term (\ref{ea})
in the effective action must arise from the gauge invariant structure
(\ref{density}). In the next sections, we will see that this term has a
very natural interpretation in terms of the matrix model formulation
of the gauge theory. Note that we expect the full one-loop 
effective action to contain
additional terms (probably an infinite series) of the form
(\ref{wact}), but all of them will be less important than
(\ref{density}) for small momenta.

\section{Review of the matrix model formulation} 

We now recall the matrix model formulation of $U(N)$ noncommutative gauge
theory, focusing on the case of pure gauge theory with maximal rank 
spatial noncommutativity ($\theta$ of rank $2p$ in $2p$ spatial dimensions).
We begin with the action for a system of bosonic D0-branes in $2p$ dimensions, 
\[
S = \int dt \; \Tr( {1 \over 2} \dot{X^i} \dot{X^i} + {1 \over 4} [X^i,X^j]^2 )
\]
For an infinite number of branes, there exists a classical solution
$X^i = x^i$ to the equations of motion such that
\[
[x^i, x^j] = i \theta^{ij} \otimes \identity_{N \times N} 
\]
For example, we may choose $\sqrt{\theta \over 2} (x^{2n-1} + i x^{2n})
= a_n \otimes \identity_{N \times N}$ where $a_n$ are matrix
representations of a set of $p$ harmonic oscillator creation
operators.\footnote{We choose $\theta$ to take the standard form 
$\theta^{ij} = \theta (i \sigma_2) \otimes \identity_{p \times p}$.} 
We may now expand the matrix theory action about this
background, defining $X^i = x^i + \theta^{ij} A_j$. We find (see e.g. 
\cite{seiberg})
\be
\label{ncg}
S = \int {d^{2p+1}x \over (2 \pi)^p {\rm Pf}(\theta)} 
\tr( {1 \over 2} \dot{A}^i \dot{A}^i-{1 \over 4} G^{km}
 G^{ln} \{F_{kl} - \theta^{-1}_{kl} \} \{ F_{mn} - \theta^{-1}_{mn} \} )
\ee
where $G^{ij} \equiv \theta^{ik} \theta^{kj}$ and $F$ is 
the
noncommutative field strength. Here, we have made the transition from
the matrix theory formalism to
the field theory formalism by considering $A$ to be a function
generated by the noncommutative coordinates $x^i$ and making the usual
substitutions
\[
[x_i, \; ] \rightarrow i \theta^{ij} \partial_j \; , \qquad \tr(f_1
\cdots f_n) \rightarrow \int {d^{2p} x \over (2 \pi)^p {\rm Pf}(\theta)} 
\tr( f_1 \star \cdots \star f_n) 
\]   
The action (\ref{ncg}) is precisely the action for pure noncommutative
gauge theory in $2p+1$ dimensions in the gauge $A_0 = 0$ (after an appropriate 
redefinition of the coordinates such that $G^{ij} \to \delta^{ij}$). We may 
ignore the $\theta^{-1}$ terms since $\int \tr (\theta^{-1} F)$
vanishes for allowed fluctuations about the perturbative
vacuum.

\section{Interpretation of the nonplanar effective action}

We would now like to interpret the term (\ref{density}) in the
nonplanar effective action in the context of the matrix model. As
reviewed in appendix B, the open Wilson line operator
$\omega(k)$ appearing in (\ref{density}) takes a very simple form in the
matrix model language, namely 
\[
\omega(k) = (2 \pi \theta)^p \; \Tr(e^{i k \cdot X}) \equiv (2 \pi
\theta)^p \rho(k) \; .
\]
This operator also has an equally simple interpretation. In type IIA
sting theory, $\rho(k)$ is precisely the leading operator coupling to $C_0(k)$,
the time component of the RR one-form \cite{tv}. Therefore, $\rho(k)$ is the
operator that measures the density of zero-brane charge. Given this
interpretation, it is natural to rewrite the effective action (\ref{ea})
in position space, and we find
\be
\label{confine}
\Gamma_{NP} = - {1 \over 2} (d-2 + N_s - {1 \over 2}{\cal N}_f) \int dt \int 
d^{2p} x \int d^{2p} y \; \rho(x,t) \rho(y,t)
|x-y| \;.
\ee
where the position space density operator is
\[
\rho(x,t) = \int {d^{2p} k \over (2 \pi)^{2p}} \; e^{-ik \cdot x} \; 
\Tr(e^{i k \cdot X(t)}) \; .
\]
Thus, the infrared singular terms in the nonplanar effective action
correspond to an instantaneous linear potential between the
D-particles making up the noncommutative space. 

This potential is exactly what one should have expected for bosonic
D-particles. Recall that the success of the BFSS matrix model in
describing nine large spatial dimensions of DLCQ M-theory depended
crucially on the presence of supersymmetries to ensure that the
classical flat directions were not lifted by quantum mechanical zero
point energies. For the purely bosonic theory, there is no
cancellation, and any widely separated pair of D-particles experience
an attractive linear potential from the string connecting
them. With an excess of fermionic degrees of freedom, the attractive
potential will be over-canceled and we will have a repulsive linear
potential between the zero-branes. 

As a check of this interpretation, we perform a direct calculation in appendix C 
of the one loop effective action in the matrix theory formulation by explicitly 
summing the zero point energies of fluctuating modes. We find that the leading 
term in the matrix theory effective action reproduces the complete expression 
(\ref{confine}) exactly, including the coefficient.     

The matrix theory calculation and the intuitive picture we have developed also 
provide checks of the sign of the potential (\ref{ea}) calculated in the field 
theory formalism. It is straightforward to verify that the negative effective 
potential in 
(\ref{ea}) found for bosonic gauge theories corresponds to an attractive 
D-particle potential,
while the positive potential occurring in theories with excess fermions
corresponds to a repulsive potential. In the next section we will
explain why the attractive potential leads to an instability while the
repulsive potential does not. 

\section{Physical consequences}

In this section, we consider the physical consequences of the potential
(\ref{confine}). At first sight, instantaneous linear forces between the 
D-particles would seem to be disastrous for the theory. One might expect that 
the 
space would simply collapse in the presence of attractive forces and blow up in 
the presence of repulsive forces. However, one must be careful to specify what 
types of fluctuations are allowed, and in particular, what boundary conditions 
to impose at infinity. 

In field theory, it is usual to require that physical fields fall off 
sufficiently rapidly at infinity. In our case, a natural condition would be to 
require that the zero-brane density remains fixed and uniform at infinity. In 
the matrix theory formalism, this should correspond to a restriction to 
fluctuation matrices $A$ of finite rank or some norm completion of this 
set.\footnote{It would be valuable to understand the appropriate restriction 
more precisely.} With such a restriction, uniform 
expansion or collapse of the space is not allowed since it alters the fields 
at infinity. We now turn to the allowed localized fluctuations.

With a fixed uniform density $\rho_0$ at infinity, conservation of zero-brane 
charge will require that $\Delta = \rho - \rho_0$ integrates to zero over the 
space for fluctuations about the perturbative vacuum, 
\[
\int d^{2p} x \Delta(x) = 0
\]
Thus, up to an infinite constant term (the energy of the initial uniform 
configuration), we may replace $\rho$ with $\Delta$ in (\ref{confine}). This 
corresponds to subtracting off the $\delta$-function term in the Wilson line 
operator (\ref{expand}).\footnote{It is obvious that the remaining terms in the 
Wilson line operator vanish for $k=0$, (i.e. $\Delta(k=0)$ = 0) so the claim 
that $\Delta(x)$ integrates to zero is justified.} The possible values of the 
resulting potential 
\[
V = C \int d^{2p}x \int d^{2p}y \Delta(x) \Delta(y) |x-y|
\]  
are more clear after a Fourier transform to momentum space, which gives 
\beas
V &=& -C (2p-1)!! \int {d^{2p} k \over (2 \pi)^p}  \Delta(k) \Delta(-k) {1 \over 
|k|^{2p+1}}\\
&=& -C (2p-1)!! \int {d^{2p} k \over (2 \pi)^p}  {|\Delta(k)|^2 \over 
|k|^{2p+1}} \; .
\eeas
In this form, it is clear that the potential is negative definite and unbounded 
below for the attractive sign ($C > 0$) and positive definite for repulsive 
D-particles ($C<0$). We may understand this as follows.  

Consider a small, local density fluctuation. By charge conservation, there must 
exist regions with positive $\Delta$ as well as regions with negative $\Delta$. 
For the repulsive D-particle forces, 
these regions will effectively attract each other with a linear potential and 
the state of 
uniform density will tend to be restored. This ``confinement'' of density 
fluctuations results in the positive definite potential we have observed.
On the other hand, with an attractive D-particle potential, regions of positive 
and negative density will tend to repel each other, and we can decrease the 
energy by an arbitrary amount by moving the 
positive and negative regions away from each other (see figure 1), or by 
increasing the magnitude of the density fluctuations. Thus, while the space is 
prevented from collapsing by boundary conditions, the theory is still unstable 
due to local bunching up of the D-particles. Nonperturbatively, the preferred 
vacuum of the matrix model is really one where all of the D-particles sit 
together in a clump, so it seems unlikely that the instability we have observed 
will terminate at any stable extended configuration of D-particles.    

The discussion in the previous paragraph ignored the tree level potential $-
[X_i, X_j]^2$. This gives rise to the $\int F_{ij} F_{ij}$ potential in the 
field theory formalism and is clearly positive definite. Physically,
it serves to suppress density fluctuations, however it is not enough
to stabilize the bosonic theory. To see this, note that since it
corresponds to a local term in the field theory action, its value for
configurations with separated positive and negative density regions
(such as the one depicted in figure 1) will be independent of the
separation of the regions. Thus, the linear decrease in energy as the
two regions are separated persists and the total potential is still
unbounded below. 

While the repulsive theories appear to be stable, they will certainly have\
some 
very unusual properties, both because the nonplanar potential is
instantaneous and because it is so strong. It would be interesting to
understand these theories better based on the D-particle intuition but
we leave this as a problem for future work.
\begin{figure}
\centerline{\epsfysize=1.4truein \epsfbox{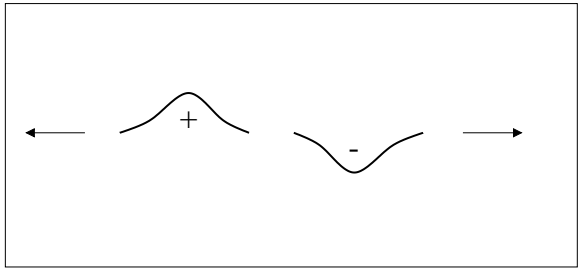}}
 \caption{Repulsion of positive and negative density fluctuations
 leads to instability of noncommutative gauge theory.}
\label{feyn}
\end{figure}
\newpage
\noindent
{\bf Supersymmetric theories}\\
\\
The D-particle intuition can also help us to understand the behavior of 
supersymmetric noncommutative gauge theories. For these theories, the
linear 
D-particle potential cancels, leaving a velocity dependent potential such as ${v^2 
\over r^3}$ or ${v^4 \over r^7}$ for maximal supersymmetry. These potentials 
will still be instantaneous, so while their effects may not be as dramatic as 
in
the nonsupersymmetric cases, they will still give rise to behavior that is 
unusual from the point of view of relativistic field theory.

In the maximally supersymmetric case, there is a nice interpretation of the 
effects of the instantaneous potential. It is well known that in the BFSS 
model, 
the leading order one-loop matrix theory potential ($v^4/r^7$ term and its 
$F^4$ 
generalization) reproduces the effects of a single supergraviton exchange in 
DLCQ eleven dimensional supergravity \cite{kt2}. In the type IIA
picture, the model 
describes D0-branes in a low-energy, non-relativistic limit, and the one-loop 
matrix model potential gives precisely the leading effects of linearized type 
IIA supergravity in the nonrelativistic limit. Thus, we expect that the leading 
nonplanar one loop effective action in maximally suspersymmetric 
$(2p+1)$-dimensional noncommutative gauge theory will reproduce the effects of 
nonrelativistic type IIA supergravity. For example, the theory will contain 
instantaneous 10-dimensional gravitational forces between any two sources of 
stress-energy.\footnote{This has been realized independently by Shiraz 
Minwalla.} We should note that the existence of gravitational forces in maximally
supersymmetric noncommutative field theories has been emphasized
previously in the context of the IKKT matrix model by \cite{aiikkt}. Similar forces 
were discovered in \cite{go} in the context of ``nonrelativistic
closed/wound string theories'' and discussed in detail in \cite{dgk}. 

A final interesting point is the claim \cite{llt} that beyond some
critical temperature, the supersymmetric theory develops a
perturbative instability similar to that in the bosonic theory. It
would be useful to understand this directly from the matrix model perspective.

\section{Comments}

In this paper, we have shown that the leading infrared singular terms in the 
nonplanar effective action of noncommutative gauge theory correspond to 
instantaneous linear potentials between the D-particles making up the 
noncommutative space. As expected, the potential is attractive with
more (adjoint) bosonic 
than fermionic degrees of freedom and repulsive with more 
fermionic degrees of freedom. In the repulsive case, the potential is positive 
definite, density fluctuations are confined, and the theory appears to be 
stable. In the attractive case, the potential is unbounded below, so the theory 
is unstable. In particular, pure noncommutative gauge theory is not a good 
quantum field theory. 

We have focused on the case of maximal rank
noncommutativity in $2p+1$ dimensions. However, a similar intuitive
picture exists in the general case of a $d+1$ dimensional
noncommutative gauge theory with $\theta$ of rank $2p$. In this case,
the lower dimensional brane picture involves a system of generalized
$(d-2p)$-branes with $2p$ transverse scalars taken in a background
which realizes the algebra of noncommutative $R^{2p}$. The leading
infrared singular terms will now correspond to instantaneous $r^{d-2p}$ potentials
between the $(d-2p)$-branes. Again, we expect the bosonic case to
be unstable and the case with excess fermions to be stable. 

The description of noncommutative gauge theories in terms of D-particles also 
offers the following useful perspective on UV-IR mixing. The number of degrees 
of freedom for a system of $N$ D-particles is proportional to $N^2$. For finite 
volume configurations with a uniform density of D-particles (for example, the 
fuzzy sphere) the number of degrees of freedom will therefore be finite and 
proportional to the volume squared. The number of degrees of freedom per unit 
volume is then proportional to the volume of the space, and only becomes 
infinite for an infinite volume distribution of D-particles. Thus, it
is clear that UV divergences (associated with having an infinite number of
degrees of freedom per unit volume) in maximal rank noncommutative field theory can
only arise for infinite volume spaces, and that this may be attributed to
the peculiar non-extensivity of the number of degrees of freedom
associated with a distribution of D-particles ($d.o.f. \propto
V^2$). UV-IR mixing in manifest, since the UV degrees of freedom involve
the large ``strings'' connecting distant D-particles (this is
consistent with previous identifications of the UV degrees of freedom
with ``bi-local fields'' \cite{ikk} or ``stretched strings'' \cite{lm2}).

Finally, we note that the contribution to the leading IR singular
terms in the 
nonplanar effective action was completely independent of the particle
masses in 
the field theory formalism. In terms of the D-particle language, adding masses 
to the matter in the theory will affect the form of the potential
between D-particles at short (or intermediate) distances but the long
distance form of the potential will remain linear. Thus, the overall
stability of the theory does not 
depend on the relative number of massless bosons and fermions, but rather on the 
relative number at all scales. Put another way, the tendency of the 
noncommutative space in its perturbative vacuum state to contract,
expand, or remain fixed as a result of the zero point energies of
fluctuating fields is determined only by the relative 
number of bosons and fermions at high energies and is independent of
particle masses. It would be interesting if similar behavior, in
particular the insensitivity to low scale supersymmetry breaking of
the long-range effects of vacuum fluctuations, exists in other theories whose
number of degrees of freedom is not proportional to volume.\footnote{We thank 
Tom Banks for
pointing out that this insensitivity does not seem to hold for theories with
supersymmetry broken by Scherk-Schwarz boundary conditions such as
those in \cite{banks}.}

\section*{Acknowledgments}
I would especially like to thank Shiraz Minwalla for valuable discussions. I 
would also like to thank Tom Banks, Nathan Seiberg, Steve Shenker, and 
Lenny Susskind for discussions and comments. I am grateful to the Rutgers High 
Energy Theory Center for hospitality while part of this work was completed. This 
work was supported in part by the Stanford Institute for Theoretical
Physics and by NSF grant 9870115.

\appendix

\section{Derivation of the quadratic effective action}

In this section, we compute the leading IR singular term in the nonplanar 
quadratic 
effective action for noncommutative gauge theory in $d$ dimensions. This 
calculation was performed in the four dimensional case in \cite{hayakawa,mst}. 

To determine the leading term we are interested in, it is enough to compute the 
contribution from a single scalar field. As shown in \cite{mst}, the gauge 
field 
and fermion contributions take precisely the same form, and have coefficients 
such that the leading term cancels in the supersymmetric case.
For calculational purposes, we work in d-dimensional Euclidean space with the 
action
\[
S = \int d^dx ({1 \over 4} F_{ab} F_{ab} + {1 \over 2} D_a \phi D_a
\phi + {1 \over 2} m^2 \phi^2)
\]
where
\[
D_a \phi = \partial_a \phi + ig [A_a, \phi]_\star \; .
\]
The effective action obtained by integrating out the scalar is given
by (star products are implied)
\[
e^{-\Gamma} = \langle e^{-\int d^d x (ig \partial_a \phi [A_a, \phi] -
{g^2 \over 2} [A_a, \phi] [A_a, \phi])} \rangle
\]
where we treat $A$ as a background field and simply perform the
gaussian integral over $\phi$. At quadratic order in $A$, we have
\beas
\Gamma &=& - {g^2 \over 2} \langle  \int d^d x (i \partial_a \phi [A_a,
\phi]) \int d^d y (i \partial_a \phi [A_a, \phi] ) \rangle \\
&& - {g^2 \over 2} \langle \int d^dx [A_a, \phi] [A_a, \phi] \rangle
\; .
\eeas
Using the momentum space propagator 
\[
\langle \phi(k) \phi(l) \rangle = (2 \pi)^d \delta^d (k+l) {1 \over
k^2 + m^2} \; ,
\]
we find that the nonplanar parts of this expectation value give
\beas
\Gamma_{NP} &=& g^2 \int {d^d k \over (2 \pi)^d} A_a(k) A_b(-k) \int 
{d^d l \over (2
\pi)^d} l_a l_b {e^{i k \times l} \over (l^2 +
m^2)((l+k)^2 + m^2)}\\
&& - g^2    \int {d^dk \over (2 \pi)^d} A_a(k) A_b(-k) \int {d^d
l \over (2 \pi)^d } {e^{i k \times l} \over (l^2 +
m^2)}\\
&=& g^2 \int {d^dk \over (2 \pi)^d} A_a(k) A_b(-k) \{-2
\partial_{\tilde{k}_a} \partial_{\tilde{k}_b} + (\partial^2_{\tilde{k}} -
m^2)\delta_{ab} \}  \int {d^d
l \over (2 \pi)^d}  {e^{i k \times l} \over (l^2 +
m^2)^2} + \dots 
\eeas
where $\tilde{k}^a = \theta^{ab} k_b$, and in the last line the dots indicate 
higher order terms arising
from expanding the propagators about $k=0$. Using
\[
{1 \over (l^2 +m^2)^2} = \int_0^\infty d \alpha \alpha e^{-\alpha(l^2 +m^2)} \; ,
\]
we may evaluate the $l$ integral, yielding
\beas
\Gamma &=& {g^2 \over (4 \pi)^{d \over 2}}  \int {d^dk \over (2 \pi)^d}
A_a(k) A_b(-k)  \{-2 \partial_{\tilde{k}_a} \partial_{\tilde{k_b}} +
(\partial^2_{\tilde{k}} - m^2)\delta_{ab} \} \int_0^\infty {d \alpha
\over \alpha^{{d \over 2} - 1}} e^{-\alpha m^2 - {\tilde{k}^2 \over 4
\alpha}} \\
&=&  {g^2 \over (4 \pi)^{d \over 2}}  \int {d^dk \over (2 \pi)^d}
A_a(k) A_b(-k)  \int_0^\infty {d \alpha
\over \alpha^{{d \over 2} - 1}} e^{-\alpha m^2 - {\tilde{k}^2 \over 4
\alpha} }(-{1 \over 2 \alpha^2} \tilde{k_a} \tilde{k_b} +
\delta_{ab}({1 \over \alpha}(1 - {d \over 2}) + {\tilde{k}^2 \over 4
\alpha^2} - m^2 ) ) \\
&=&  -{g^2 \over 2 \cdot (4 \pi)^{d \over 2}}  \int {d^dk \over (2 \pi)^d}
A_a(k) A_b(-k) \tilde{k}_a \tilde{k}_b \int_0^\infty {d \alpha
\over \alpha^{{d \over 2} + 1}} e^{-\alpha m^2 - {\tilde{k}^2 \over 4
\alpha} }\\
&&+ {g^2 \over (4 \pi)^{d \over 2}}  \int {d^dk \over (2 \pi)^d}
A_a(k) A_a(-k)  \int_0^\infty d \alpha \partial_{\alpha} \left( {1
\over \alpha^{{d \over 2} - 1}} e^{-\alpha m^2 - {\tilde{k}^2 \over 4
\alpha} } \right) \\
\eeas
The vanishing of the last line provides a check of the calculation so
far since the $A_a A_a$ structure is not consistent with gauge
invariance. For small values of momenta (or setting $m$ to zero, the 
remaining term gives our final result,
\[
\Gamma_{NP} = \int {d^d k \over (2 \pi)^d} A_a(k) A_b(-k) \left( - {g^2
\over 2 \pi^{d \over 2} } \Gamma({d \over 2}) {\tilde{k}^a
\tilde{k}^b \over |\tilde{k}|^d} \right)
\]
Including general adjoint matter content and translating to Minkowski
space (by switching the sign), we recover the expression
(\ref{ea}) stated in section 2.   

\section{Open wilson lines in the matrix theory language}

For completeness, we recall here the derivation that the simplest open
Wilson line operator in the field theory formalism corresponds to the
zero-brane density operator in the matrix theory formalism
\cite{iikk}. It will be
useful to recall that 
\[
e^{i k \cdot x} \star f(x) \star e^{-i k \cdot x} = f(x + \tilde{k})
\]
where $\tilde{k}^a = \theta^{ab} k_b$ as usual. Then starting in the
field theory formalism, we have
\beas
\omega(k) &=& \int d^dx e^{i k \cdot x} \tr \left( P_\star e^{ig \int_0^1 
A_a(x + \sigma \tilde{k}) \tilde{k}^a d \sigma} \right)\\
&=& \lim_{N \to \infty}  \int d^dx \tr \left( P_\star e^{
\sum_{n=1}^\infty {ig \over N} \int_0^1 A_a(x + {n \over N} \tilde{k}) 
\tilde{k}^a } \right) \star e^{i k
\cdot x}\\
&=&  \lim_{N \to \infty} \int d^dx  \tr \left( \prod_\star e^{{ig
\over N}  A_a(x + {n \over N} \tilde{k}) \tilde{k}^a}  \right) \star e^{i k
\cdot x}\\
&=&  \lim_{N \to \infty} \int d^dx  \tr \left( \prod_\star e^{ik \cdot
x {n\over N}} \star e^{{ig \over N}  A_a(x) \tilde{k}^a} \star e^{-i k
\cdot x {n \over N}}  \right)  \star e^{i k \cdot x}\\
&=&  \lim_{N \to \infty} \int d^dx  \tr \left( e^{ik \cdot
x \over N} \star e^{{ig \over N}  A_a(x) \tilde{k}^a} \right)^N\\
&=&  \lim_{N \to \infty} \int d^dx  \tr \left( e^{{1 \over N} ik_a (x^a + g
\theta^{ab} A_b)} \right)^N\\
&\to&  (2 \pi \theta)^p \Tr \left( e^{i k \cdot X} \right)
\eeas
where the trace in the last line denotes the full operator
trace. Gauge invariance is manifest in the matrix model formalism,
where the noncommutative gauge transformations are simply unitary
transformations $X^i \to U X^i U^{-1}$. 

\section{Direct calculation of the effective action from matrix theory}
 
In this section, we verify the suggestion (\ref{density}) for the
dominant gauge invariant 
structure in the nonplanar effective action by a direct calculation in Matrix 
Theory. We have argued that (\ref{confine}) is consistent with the expectation that there 
should be a linear potential between D0-branes in nonsupersymmetric matrix 
theory. This expectation arises from calculations performed for widely separated 
D-particles in a diagonal background, whereas we are presently
considering 
D-particles in a noncommuting configuration with uniform D-particle charge spread 
over $2p$ spatial dimensions. It is therefore important to check directly that 
the linear potential (\ref{confine}) does arise in this background.

To determine the one loop effective action in the matrix quantum mechanics, we 
simply need to sum the zero point energies of all fluctuating modes expanded 
about the background we are interested in. In practice, it is again simplest to 
consider the theory with an additional scalar field and then integrate out the 
scalar. Thus, we start with the lagrangian
\[
{\cal L} = \Tr( {1 \over 2} \dot{X}^i \dot{X}^i + {1 \over 4} [X^i, X^j]^2 + {1 \over 
2} \dot{\Phi} \dot{\Phi} + {1 \over 2} [X^i, \Phi]^2 ) \; .
\]
The mass squared matrix for $\Phi$ may be written
\[
M^2 = (X^i \otimes \identity -  \identity \otimes X^i)^2
\]
The effective potential is then given by the sum of zero point energies,
\bea
V &=& {1 \over 2} \Tr(\sqrt{M^2}) \nonumber \\
&=& {1 \over 2} \Tr(\sqrt{(X^i \otimes \identity -  \identity \otimes X^i)^2}) 
\nonumber \\
&=& -{1 \over 2 \pi^{2p+1 \over 2}} \Gamma({2p+1 \over 2}) \int d^{2p} k 
\; \Tr(e^{i 
k_i (X^i \otimes 1 -  1 \otimes X^i)}) {1 \over |k|^{2p+1}} + 
{\rm commutators} \nonumber \\
&=& -{1 \over 2 \pi^{d \over 2}} \Gamma({d \over 2}) \int d^{d-1} k \; \Tr(e^{i k 
\cdot X}) \Tr(e^{- i k \cdot X})  {1 \over |k|^d}+ {\rm commutators}
\label{mtresult}
\eea
To see that these manipulations are sensible, note that for the background $X^i 
= x^i$, the matrix $M^2$ is diagonal and all of the eigenvalues are positive. 
Thus, for $X^i = x^i + \theta^{ij} A_j$, there is a well defined expansion of 
the square root in the second line powers of $A$. The Fourier transformed 
expressions in the third and fourth line also admit well defined expansions in 
powers of $A$ (see appendix B), however these imply a fully symmetrized ordering 
prescription (Weyl ordering) while the expression in the second line is not 
fully symmetrized. As we have indicated, the difference will be terms involving 
commutators of $X^i$'s and these will correspond in the field theory formalism 
to higher order terms of the form (\ref{wact}) in the effective
action involving field 
strengths and covariant derivatives of field strengths inserted in the Wilson 
line.

It is straightforward to repeat this calculation for gauge field fluctuations 
and fermions (it is essentially identical to the Matrix theory effective action 
calculations performed in \cite{kt}). In each case, the leading order potential 
is identical to (\ref{mtresult}) with additional factors of $(d-2)$ and 
$-{1\over 2} {\cal N}_f$ respectively. Thus, for the general case, we may write 
the leading 
order effective potential in the matrix theory formalism as
\[
V = -{1 \over 2 \pi^{d \over 2} }\Gamma({d \over 2}) (d-2 + N_s -
{\cal N}_f) \int d^{d-1} k \rho(k) \rho(-k) {1 \over |k|^d} \; .
\]
After a Fourier transform to position space, this gives precisely
the expression (\ref{confine}) that we deduced from the field 
theory result.

\bibliographystyle{plain}


\end{document}